\documentclass[pre, reprint, superscriptaddress, amsmath, amssymb, floatfix]{revtex4-2}
\usepackage{kotex}
\usepackage[dvipsnames]{xcolor}
\usepackage{float}
\usepackage{graphicx}
\usepackage{dcolumn}
\usepackage{bm}
\usepackage[normalem]{ulem} 

\usepackage{cleveref}
\usepackage[protrusion=true,expansion=true]{microtype}
\usepackage{multirow}
\usepackage{makecell}
\usepackage[ruled,vlined,linesnumbered]{algorithm2e}

\begin{document}

\title{AI-Driven Stabilization in Power Grids through Controlling Line Admittances}

\author{Sangjoon Park}
\affiliation{CCSS, KI for Grid Modernization,
Korea Institute of Energy Technology, Naju, Jeonnam 58330, Korea}

\author{Hoyun Choi}
\affiliation{School of Computational Sciences, Korea Institute for Advanced Study, Seoul, 02455, Korea}

\author{Yongsun Lee}
\affiliation{CTP and Department of Physics and Astronomy, Seoul National University, Seoul 08826, Korea}

\author{Seungchan Jo}
\affiliation{Department of Electrical and Computer Engineering, Seoul National University, Seoul 08826, Korea}

\author{J\"urgen Kurths}
\affiliation{Potsdam Institute for Climate Impact Research, Telegraphenberg, D-14415, Potsdam, Germany}

\author{B. Kahng}
\affiliation{CCSS, KI for Grid Modernization,
Korea Institute of Energy Technology, Naju, Jeonnam 58330, Korea}
\date{\today}

\begin{abstract}
The global transition from traditional power plants to renewable energy sources introduces new challenges in grid stability, primarily because inverter-based technologies provide insufficient inertia. To address this, we introduce an artificial intelligence algorithm that autonomously stabilizes power grids by adaptively tuning admittance regulators in response to disturbances. This Adaptive Admittance Controller (AAC) algorithm not only stabilizes the system in real time but also identifies the best regulator locations, thereby unifying grid planning and real time control within a single framework. When tested on a real UK power grid, the AAC markedly reduces frequency deviations and rapidly restores nominal operation. In addition, the algorithm isolates a small number of key regulators and intervenes only on these, lowering both system complexity and cost. The AAC algorithm further reduces the nonlinearity effect, quickly stabilizing the frequency and power flow. This intelligent control scheme enables power grids to reliably return to stable operating conditions under a broad spectrum of fault scenarios. The proposed framework can also be used to mitigate cascading failures by adaptively controlling critical links in a variety of networked infrastructures, such as cascades of traffic congestion on road networks or fuse failures in energy-saving systems.

\end{abstract}

\maketitle

\section{introduction}
Electrical power grids underpin modern society, enabling almost every aspect of daily life—from hospitals and data centers to residential and industrial operations~\cite{obama2013presidential,rinaldi2001identifying}. The worldwide move toward carbon-neutral energy systems represents one of the most ambitious infrastructure overhauls in human history. Yet this transformation also brings significant vulnerabilities: as renewable resources replace traditional generators, power systems lose the rotational inertia that has historically damped disturbances, endangering the stability that billions of people rely on~\cite{shaukat2018survey,gulraiz2025advancing,crivellaro2020beyond,khalid2024smart}.

The replacement of synchronous generators with inverter-based resources diminishes overall system inertia, making the grid more vulnerable to frequency disturbances. At the same time, the intrinsic variability of renewable sources introduces a level of operational complexity not previously encountered~\cite{yap2019virtual,kerdphol2018virtual, smith2022effect}. These challenges become even more pronounced during transmission line outages, where the loss of a single line can set off cascading instabilities throughout interconnected networks.

Recent large-scale blackouts illustrate the severe impact of grid instability. The 2019 UK blackout~\cite{bialek2020does} and the 2021 Texas power crisis~\cite{zhang2022texas, flores20232021} were both triggered by transmission line outages caused by extreme weather, putting millions of people at risk and resulting in economic losses of billions of dollars. These incidents underscore the urgent demand for new stability management approaches that can function reliably in grids with a high share of renewable generation~\cite{sharma2021major, raza2022analysis,sturmer2024increasing}.

Frequency is a primary real-time metric for assessing the balance between power supply and demand~\cite{pagnier2019inertia,fernandez2022frequency}. Under normal conditions, it is carefully maintained within tight bounds; nonetheless, transmission line faults can trigger abrupt deviations that endanger both global and dynamic stability~\cite{schafer2018dynamically}. Rapid suppression of these frequency excursions is crucial to prevent cascading failures, which may lead to large-scale blackouts and severe economic as well as social consequences~\cite{pahwa2014abruptness,zhang2016optimizing,dai2022risk}. The evolution of system frequency is typically described by the swing equation~\cite{alexander1986oscillatory,qiu2020swing}.

Traditionally, stability has been improved by modifying the infrastructure, either by constructing additional transmission lines or upgrading existing ones~\cite{odor2024improving, schafer2022understanding}. Yet, these approaches are increasingly hindered by public opposition, substantial costs, protracted permitting procedures, and environmental concerns~\cite{cain2013drives, cohen2016empirical}. In recent years, numerous works have investigated boosting grid stability through flexible alternating current transmission system devices, such as thyristor-controlled series capacitors~\cite{lee2016integrating, AZIMI_SHAHGHOLIAN_2019, nkan2023enhancement}, which regulate the admittance of transmission lines. Their overall benefit, however, is constrained by two tightly linked issues: determining where to install them (a planning task) and how to tune their admittances in operation (an operational task). The enormous combinatorial space of possible regulator deployments and settings makes conventional optimization techniques computationally demanding, and poorly chosen configurations may even worsen stability.

These challenges align well with reinforcement learning (RL), which is adept at handling high-dimensional optimization problems in dynamical systems. RL has achieved notable success in a wide range of autonomous systems by enabling consistent and adaptive control strategies~\cite{silver2016mastering,mirhoseini2021graph,degrave2022magnetic,jo2024self,jacob2024real}. Within power grid applications, RL-based methods have shown promise for tasks such as power dispatch—i.e., regulating power supply levels~\cite{lee2024reinforcement}—and for admittance control of individual transmission lines~\cite{huang2022damping, ernst2004power}.

Such studies, however, treat planning and operation as distinct stages: they first determine regulator placements on inter-area tie-lines and only then design control schemes for those predetermined sites. Although this decoupled methodology is computationally tractable, it is fundamentally suboptimal. Once regulator locations are fixed, it is impossible to realize truly optimal control for faults arising at different points in the system.

To this end, we propose an Adaptive Admittance Controller (AAC) algorithm based on an AI-driven approach, in which RL combined with a graph neural network (GNN) autonomously learns a control policy directly from interaction with the power grid environment. Once trained, the algorithm can instantly determine which lines to control and by how much for any given fault scenario within a single calculation of the GNN. By integrating both planning and real-time operation, the RL-based algorithm demonstrates two key capabilities: (i) performs intelligent, selective intervention by strongly mitigating high-risk contingencies while avoiding superfluous actions during minor disturbances; (ii) determines the minimum necessary subset of transmission lines on which to install regulators to achieve near-optimal control, thereby substantially reducing deployment costs. Evaluating more than 105 single-line fault scenarios in a simplified UK national grid model~\cite{pagnier2019inertia}, our approach provides a computationally efficient and economically viable solution, supporting the secure operation of future power systems with high penetrations of renewable energy sources.

\section{Frequency Stabilization and Essential Placements}

We employ two power grids: the UK power grid with a reduced number of buses and lines~\cite{pagnier2019inertia} and a synthetic so-called SHK grid~\cite{schultz2014random}. In the main text, we present results for the UK power grid, while those for SHK are in the Supplementary Information (SI).

\subsection{Optimal reduction of frequency fluctuations}
The frequency dynamics of the power grid is described by the swing equation~\cite{alexander1986oscillatory,qiu2020swing}:
\begin{equation}
m_{i}\ddot{\theta}_{i}+\gamma_{i} \dot{\theta}_{i}= P_{i}+ \sum_{j \neq i}{K_{ij}\sin(\theta_{j}-\theta_{i})},
\label{eq:swing}
\end{equation}
where $m_i$ represents the inertia of the turbine in the bus $i=\{1,\dots,N\}$. Note that $m_i$ in the UK grid is highly heterogeneous due to the composition of various generators. $\theta_i$ is the phase of oscillator $i$ and $\dot{\theta}_i$ is the angular velocity of the oscillator $i$, defined with respect to the rotating reference frame with the standard  frequency (e.g., 50 Hz in the UK). So, $\dot{\theta}_i = 0$ in a stable state.; $\gamma_i$ is the damping coefficient; and $P_i$ is the power ($P_i>0$ for generators and $P_i<0$ for consumers), which must fulfill the balance condition $\sum_i P_i=0$. The coupling strength is composed of $K_{ij}=V_i V_j Y_{ij}$, where $V_i$ and $V_j$ are the voltages of the buses $i$ and $j$, which are set to 1 under the assumption that the system is limited to the high voltage AC level, and $Y_{ij}$ is the admittance of the transmission line between the buses $i$ and $j$. $Y_{ij}=0$ if buses $i$ and $j$ are not connected. The UK power grid system we consider is composed of $N=54$ and the total number of transmission lines $L$ is 114, which is reduced using the Kron reduction.

In this study, we focus on single-line fault scenarios, which are among the most common disturbances in real power systems and therefore provide a natural starting point for validating the control framework. To simulate transmission line fault between bus $u$ and $v$, we set $Y_{uv} = K_{uv}=0$. Due to the heterogeneity of buses, the complex grid topology, and their nonlinear dynamics [Eq.~\eqref{eq:swing}], each line fault results in highly disparate responses across the entire system. Therefore, we consider all single-line fault scenarios that do not disconnect the grid, and specifically, 105 lines for the UK grid.

\begin{figure*}
\centering
\includegraphics[width=0.99\linewidth]{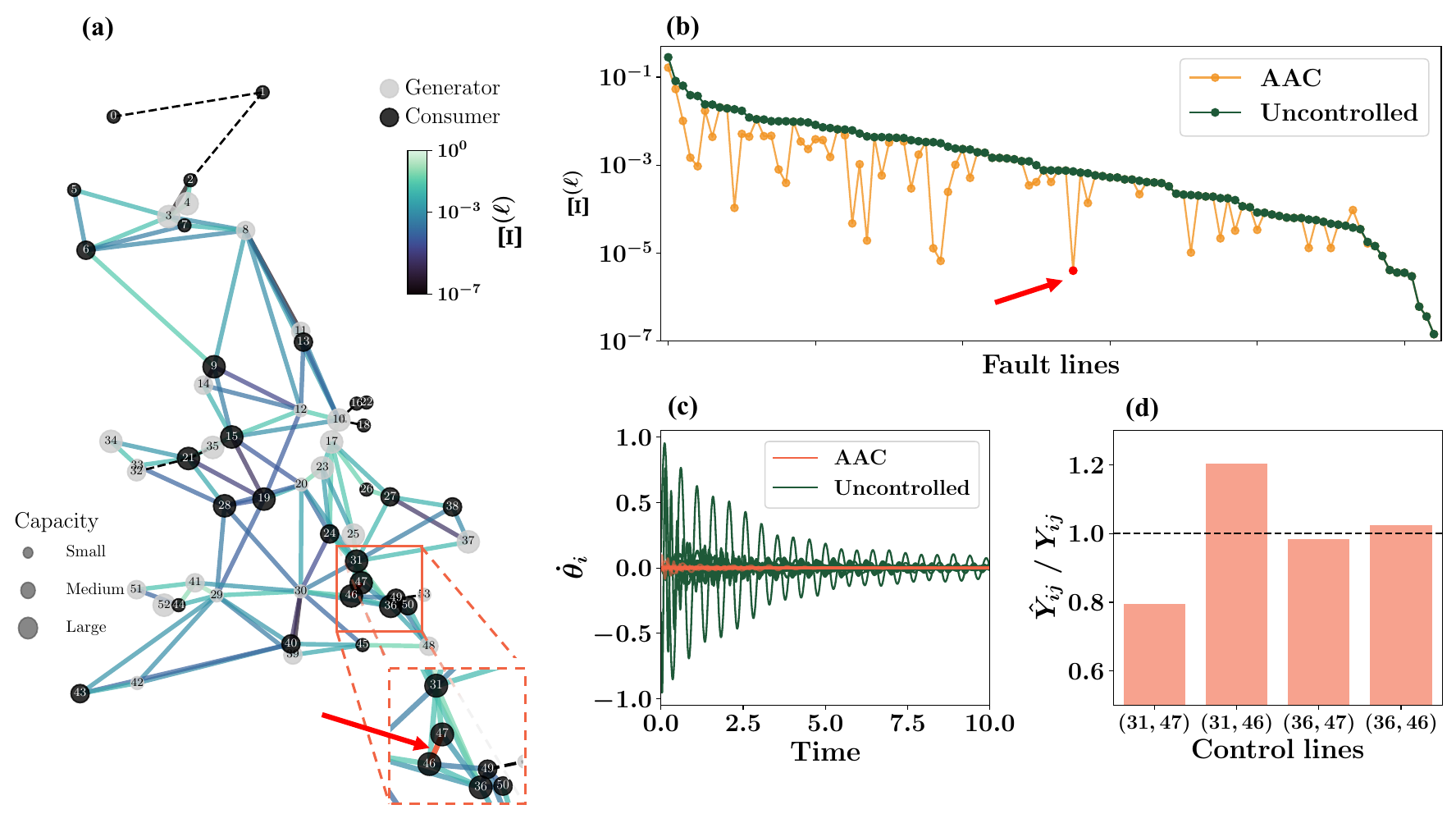}
\caption{Performance of the AAC algorithm under single-line faults in the UK power grid. (a) The color on each line ($u,v$) indicates $\Xi^{(\ell)}$ in Eq.~\eqref{eq:xi}. Node size indicates the absolute value of the power supply $|P_i|$ while generators are marked in gray and consumers in black. Dashed black lines indicate the lines excluded from our consideration as they disconnect the grid. (b) Comparison of the performance of the AAC algorithm for each fault case: controlled fluctuations (orange) vs. the uncontrolled case (green). In (a) and (b), the red arrows indicate the line between buses 46 and 47, selected as a well-behaved line for our studies, and exhibit the responses in (c) and (d).
(c) Time evolution of the angular velocities $\dot{\theta}_i$, showing that the AAC algorithm (orange) rapidly suppresses the system's fluctuations.
(d) The ratio of the adjusted admittance $\hat{Y}_{ij}$ by the AAC algorithm to the original admittance $Y_{ij}$.}
\label{fig:1}
\end{figure*}

The frequency fluctuation triggered by a fault in line $\ell=(u,v)$ is measured as $\Xi^{(\ell)}$~\cite{lee2024reinforcement}:
\begin{widetext}
\begin{equation}
\Xi^{(\ell)} \equiv \frac{1}{T} \int_0^T dt \Bigg[
\frac{1}{\sum_i m_i} \sum_i m_i \dot{\theta}_i^2(t)
- \left( \frac{1}{\sum_i m_i} \sum_i m_i \dot{\theta}_i(t) \right)^2
\Bigg],
\label{eq:xi}
\end{equation}
\end{widetext}
where $T$ is the measurement time interval, which is set to 10 seconds. This time window matches the usual interval before system operators issue new dispatch commands to restore stability, enabling the assessment of frequency fluctuations during the period preceding these corrective measures.
$\Xi^{(\ell)}$ can be understood as inertia-weighted frequency fluctuations as a consequence of the line fault $\ell$. Fig.~\ref{fig:1}(a) shows $\Xi^{(\ell)}$ for each line fault on a logarithmic scale. The scale of $\Xi^{(\ell)}$ varies widely among faults: the largest value is about $10^7$ times larger than the smallest. This wide range of heterogeneity highlights a key challenge in distinguishing between faults that require intervention from those where intervention may not be necessary or even harmful.

The AAC algorithm addresses this challenge through a selective intervention. It consists of a graph neural network (GNN) designed to output an admittance adjustment for all transmission lines based on the power grid state and line fault. For the detailed implementation of AAC algorithm, see Sec.~\ref{sec:method}. Fig.~\ref {fig:1}(b) compares $\Xi^{(\ell)}$ for each line fault, with (orange) and without (green) adjustment applied via the AAC algorithm. In the plot, the $x$-axis represents all single-line faults in the UK power grid, ordered by the uncontrolled values of $\Xi^{(\ell)}$, such that higher impact faults appear on the left. In the range where $\Xi^{(\ell)}$ is large, the AAC algorithm substantially reduces its value, whereas in low-impact regions, the values are nearly identical. This indicates that the AAC algorithm effectively suppresses frequency fluctuations in severe faults, while avoiding non-necessary intervention in minor faults. Although one low-risk scenario shows a slight increase in $\Xi^{(\ell)}$, highlighting the importance of deciding whether to intervene, its magnitude remains far below high-risk levels and does not pose a threat to overall stability. On average, it decreases $\Xi^{(\ell)}$ by approximately 53\%.

Figs.~\ref{fig:1}(c) and (d) illustrate the cases marked in a red-filled circle in Fig.~\ref{fig:1}(b), which corresponds to the line fault between buses 46 and 47, marked with an arrow in Fig.~\ref {fig:1}(a). Fig.~\ref{fig:1}(c) shows that after applying the AAC algorithm, the oscillations are drastically suppressed in the amplitudes of the angular velocities compared to those in the uncontrolled case. Fig.~\ref{fig:1}(d) presents the adjustments in the selected line admittance. It reveals that stabilization requires control of only a few lines and often involves both increasing and decreasing admittance.

In summary, the AAC algorithm efficiently reduces frequency fluctuations, particularly for lines with high volatility. In contrast, it avoids superfluous intervention for lines with low volatility, where its effect is negligible. This targeted and resilient approach is crucial for practical applications, as it enhances system stability and reduces unnecessary control actions and resource utilization.

\begin{figure*}[ht]
\centering
\includegraphics[width=0.99\linewidth]{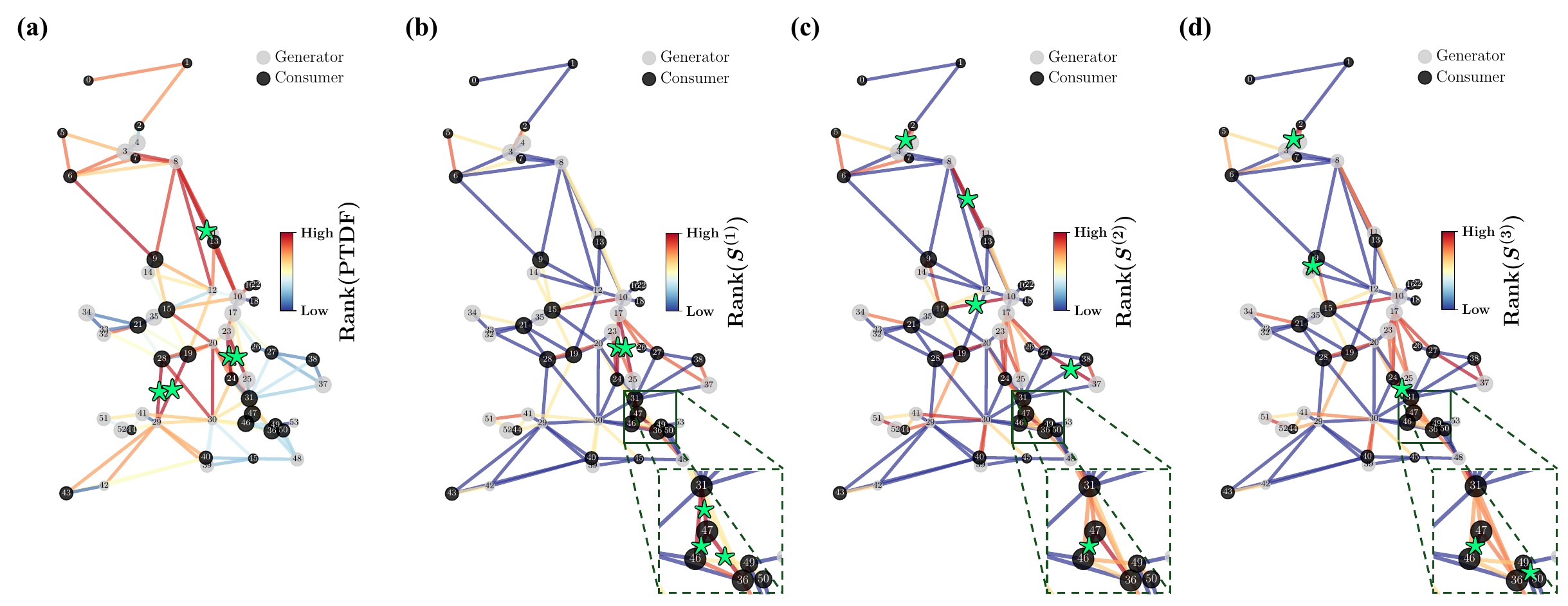}
\caption{Rankings of transmission lines for each metric, (a) the PTDF; (b) $S^{(1)}$; (c) $S^{(2)}$; and (d) $S^{(3)}$. The top five-ranked lines for each metric are marked with green stars. The significant differences between the top five-ranked lines selected by the traditional method (a) and by the AAC algorithm (b-d) demonstrate the importance of using control-aware criteria to identify critical regulator locations.
}
\label{fig:2}
\end{figure*}

\subsection{Optimal placement of regulators}
\begin{figure}[ht]
\centering
\includegraphics[width=0.99\linewidth]{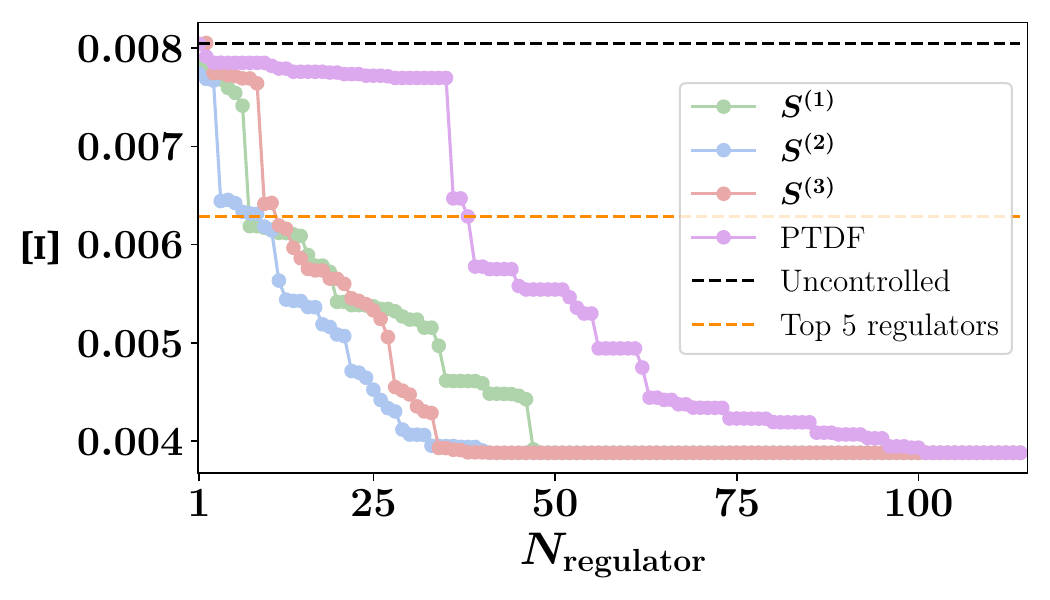}
\caption{Averaged frequency fluctuation $\Xi$ vs. the number of regulators $N_{\rm regulator}$ installed on the transmission lines in the UK power grid. The regulators are ordered by rank for each metric. $S^{(2)}$ achieves the best performance by only using less than one-third of all regulators. The orange dashed line is obtained from the AAC algorithm trained only with the top five ranked regulators in $S^{(2)}$.
}
\label{fig:3}
\end{figure}

Although the AAC algorithm is trained in an environment where it can control the admittance of all non-failed lines, in practice it is observed to control only a few lines. Using its behavior, we can reduce the number of admittance regulators in advance, enabling cost-effective operation.

Previous research in this regard used the power transfer distribution factor (PTDF) approach~\cite{wood2013power}, which focuses on how the current flow in each line responds to a change in the power of a specific bus. This approach does not account for the case where the admittance of a particular line is changed. 
To address this issue, we introduce three ranking metrics for line ($i,j$), based on the average performance of the AAC algorithm under different fault scenarios, and we assess their effectiveness in comparison with the conventional PTDF-based ranking.

\begin{itemize}
    \item[$S^{(1)}$]: The number of scenarios where the AAC algorithm controls line $(i, j)$.
    \item[$S^{(2)}$]: The ratio of admittance changed by the AAC algorithm to the original value, $\hat{Y}_{ij}/Y_{ij}$.
    \item[$S^{(3)}$]: The difference in admittance changed by the AAC algorithm to the original value, $|\hat{Y}_{ij} - Y_{ij}|$.
\end{itemize}

In Fig.\ref{fig:2}(a)$-$(d), we mark the rank in color on each line for each measure, PTDF, $S^{(1)}$, $S^{(2)}$, and $S^{(3)}$, respectively. Lines with the top five ranks are marked with a star on their lines.

To evaluate the effectiveness of each ranking under the AAC algorithm, we employ average frequency fluctuations: $\Xi \equiv {1 \over L_{\rm f}} \sum_{\ell}{\Xi^{(\ell)}}$, where $L_{\rm f}$ denotes the number of fault scenarios. We then measure $\Xi$ while decreasing the number of admittance regulators $N_{\rm regulator}$ for each metric. Fig.~\ref{fig:3} illustrates $\Xi$ for each measure as a function of $N_{\rm regulator}$. We reach the minimum value of the fluctuations with $\approx 0.004$ by installing regulators on only 35$-$45 top-rank lines out of the total $114$ possible lines. In comparison, the PTDF-based ranking needs around 100 regulators to achieve a similar minimum.
In particular, $S^{(2)}$ is found to be the best strategy, in which the minimum fluctuation level can be reached with 35 regulators and provides the lowest $\Xi$ with fewer regulators.

Fig.~\ref{fig:3} also shows that installing only five regulators in the order of $S^{(2)}$ significantly drops $\Xi$, indicating the existence of a sweet spot between cost and grid stability. Thus, we argue that installing five regulators with the highest ranks would efficiently reduce fluctuations and be cost-effective.

To verify the stabilization performance under the cost-effective regulators placement, we again train the AAC algorithm using only the top five regulators in $S^{(2)}$. As marked with an orange dashed line, the resulting performance is slightly better than that of the original model, which is trained with all regulators (except for the fault line) and then restricted to use only the five.

\begin{figure}
\centering
\includegraphics[width=0.99\linewidth]{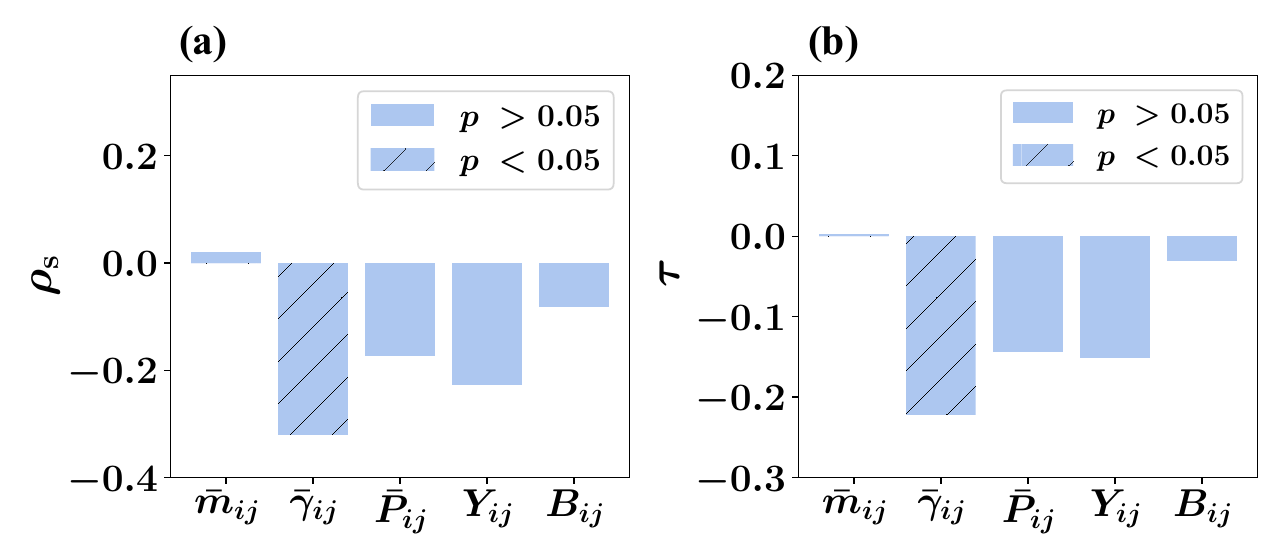}
\caption{Rank correlation between $S^{(2)}$ and various physical quantities of each line: average inertia $\bar{m}_{ij}$, average damping coefficient $\bar{\gamma}_{ij}$, average power $\bar{P}_{ij}$, admittance $Y_{ij}$, and edge betweenness centrality $B_{ij}$, where $\overline{(\cdot)}$ denotes the average over the two end nodes of the line. (a) Spearman rank correlation coefficient $\rho_s$ and (b) Kendall's tau correlation coefficient $\tau$. Filled bars indicate statistical significance ($p < 0.05$) and open bars indicate non-significance ($p > 0.05$). Lines that were never controlled by the AAC algorithm across all fault scenarios are excluded. Only $\bar{\gamma}_{ij}$ shows a statistically significant, albeit weak, negative correlation with $S^{(2)}$ in both measures.
}
\label{fig:R1}
\end{figure}

We further analyzed how the best-performing metric $S^{(2)}$ relates to different grid characteristics. In particular, we computed the Spearman rank correlation coefficient $\rho_s$ and Kendall's tau correlation coefficient $\tau$ between the ranking of $S^{(2)}$ and the rankings of several node- and line-based quantities: the average inertia $\bar{m}_{ij}$, average damping coefficient $\bar{\gamma}_{ij}$, average power $\bar{P}_{ij}$, admittance $Y_{ij}$, and edge betweenness centrality $B_{ij}$. Here, $\overline{(\cdot)}$ denotes the average over the two terminal nodes of line $(i, j)$. As illustrated in Fig.~\ref{fig:R1}, most of these correlations are weak and not statistically significant ($p > 0.05$). The only statistically significant correlation ($p < 0.05$) appears for $\bar{\gamma}_{ij}$ in both correlation measures, indicating that the AAC algorithm generally intervenes more strongly on lines whose incident buses have lower average damping coefficients. Nevertheless, no single physical variable shows a correlation strong enough to fully account for the observed ranking structure. This is because the control actions selected by the AAC algorithm embody the combined influence of multiple, interdependent factors arising from the nonlinear dynamics in Eq.~\eqref{eq:swing}.

In short, the AAC algorithm offers a unified framework that not only determines the optimal line on which to place the admittance regulator, but also ensures efficient system operation. Moreover, the ranking produced by the AAC algorithm surpasses that of the traditional PTDF-based method and reflects intricate, nonlinear dependencies among grid characteristics that cannot be attributed to any single physical parameter.

\section{Connecting transient behavior with the steady state}
We have shown that the AAC algorithm suppresses frequency fluctuations in the transient regime. Nevertheless, stability in the steady state is not guaranteed due to the nonlinearity of Eq.~\eqref{eq:swing} and the heterogeneity of the power grid.
To evaluate steady state stability, we examine the phase of each bus with and without the AAC intervention. The phase space is defined as a set of phases of each oscillator $\{\theta_1,\cdots,\theta_N\}$. Suppose that the system is in steady state with $\{\theta^o_1,\cdots,\theta^o_N\}$. A single line fault causes a disturbance in the phase space, and then the system moves to another steady state $\{\theta^*_1,\cdots,\theta^*_N\}$, where $\dot{\theta}^*_{i}=0$ for all $i$.

In the steady state, we measure the fraction of power flow relative to the capacity of each line as follows:
\begin{equation}
    I_{ij} \equiv \left|\sin (\theta_i - \theta_j)\right|.
\end{equation}

Changes in steady state lead to large variations in $I_{ij}$, which may introduce instability to the system. Therefore, if the grid is optimally controlled in response to a disturbance, $I_{ij}$ should return to its original value once steady state is recovered. We define the power flow variation $d^{(\ell)}$ in the phase space as
\begin{equation}
d^{(\ell)}=\frac{1}{L-1} \sum_{(i,j)} \left| I_{ij}^{o} - I_{ij}^{(\ell)} \right|,
\label{eq:distance}
\end{equation}
where $I^{o}_{ij}$ represents $I_{ij}$ of line $(i,j)$ in the original steady state and the superscript $(\ell)$ denotes the $\ell$-th line fault scenario.

\begin{figure}
\centering
\includegraphics[width=0.99\linewidth]{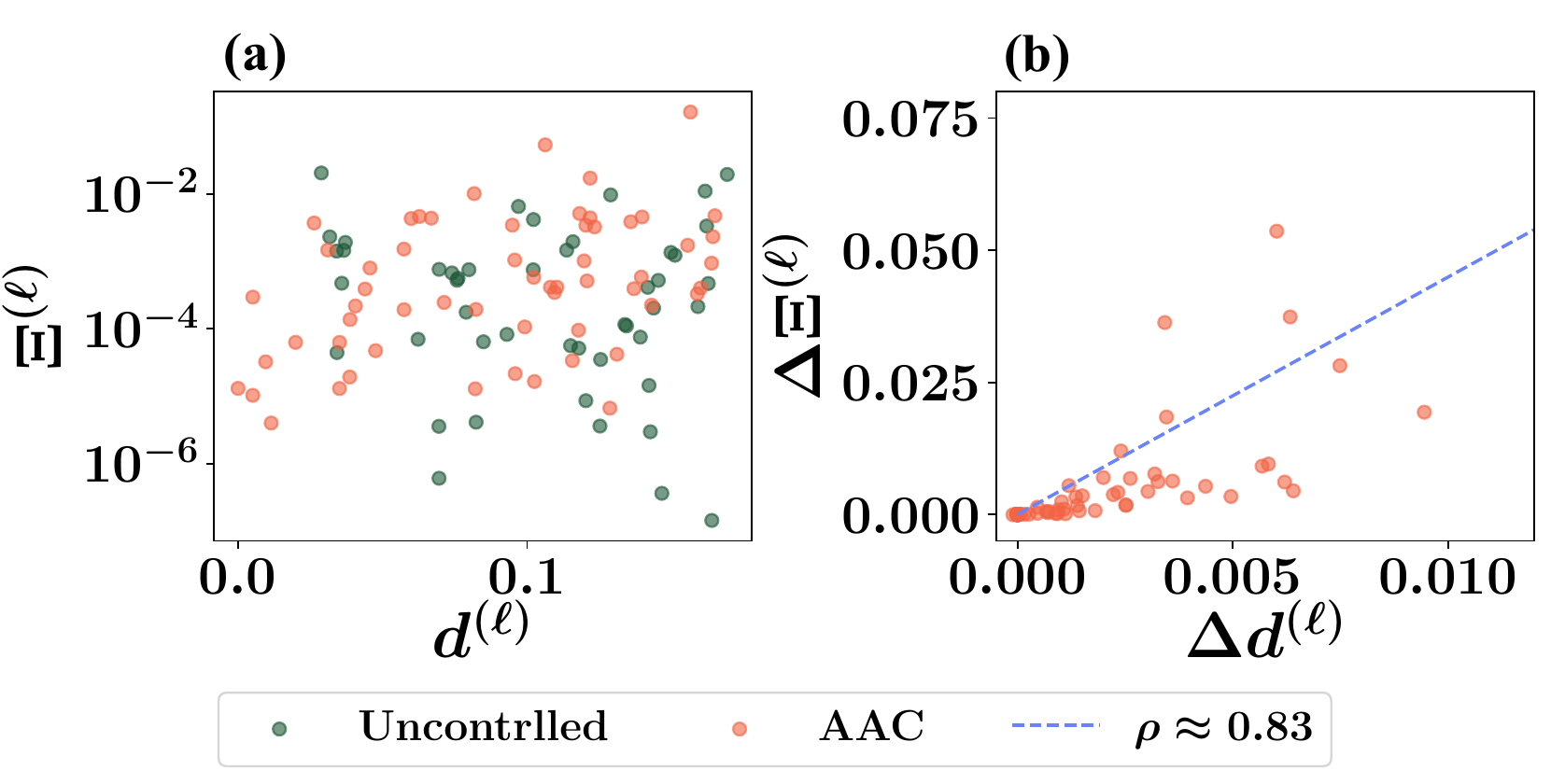}
\caption{Phase space analysis of the AAC algorithm's operating mechanism. (a) Scatter plot showing the relationship between frequency fluctuations $\Xi^{(\ell)}$ and power flow variation $d^{(\ell)}$. The two quantities exhibit largely independent behavior. (b) The scatter plot shows the relationship between the reduction in frequency fluctuations $\Delta \Xi^{(\ell)}$ and the power flow restoration $\Delta d^{(\ell)}$. Each point corresponds to $\ell$-th single-line fault in the UK grid.
}
\label{fig:4}
\end{figure}

We first investigate the relationship between the transient regime frequency fluctuations $\Xi^{(\ell)}$ and the deviation of the power flow $d^{(\ell)}$ in the steady state. As shown in Fig.~\ref{fig:4}(a), these quantities exhibit negligible correlation over different $\ell$ regardless of the adjustment of the AAC algorithm. This reflects the weak dependency between the two regimes due to the inherent nonlinearity and heterogeneity of the system.

In contrast, Fig.~\ref{fig:4}(b) presents a markedly different result when examining control-induced changes. We first observe that $d^{(\ell)}$ decreases after AAC intervention for all $\ell$, corresponding to an increase in the power-flow restoration measure $\Delta d^{(\ell)}$, demonstrating that AAC stabilizes both transient dynamics and steady state conditions. Furthermore, the Pearson correlation coefficient $\rho$ between the decrease in frequency fluctuation $\Delta \Xi^{(\ell)}$ and the restoration of power flow $\Delta d^{(\ell)}$ is close to one ($\rho \approx 0.83$), indicating that the AAC algorithm effectively relates the two regimes.

\section{Discussion}
We illustrate that artificial intelligence (AI) is capable of effectively stabilizing the power grid by strategically and adaptively controlling transmission line admittances in response to single-line fault events, which is the most common type of disturbance. The proposed Adaptive Admittance Controller (AAC) algorithm integrates the typically distinct layers of grid planning and real-time operation into a single reinforcement learning framework.
We address a few key issues faced by grid operators transitioning to renewable grids.
First, the AAC algorithm effectively mitigates power grid instability, achieving an average 53\% reduction in frequency fluctuations while avoiding unnecessary interventions during negligible disturbances (Fig.~\ref{fig:1}).
Second, the AAC-derived ranking metric $S^{(2)}$ enables optimal regulator placement, requiring significantly fewer regulators than the conventional PTDF-based approach.
Specifically, placing regulators on just five critical transmission lines achieves near-optimal stabilization while reducing implementation costs by more than 95\% compared to comprehensive deployment (Figs.~\ref{fig:2} and \ref{fig:3}).
Together, these capabilities directly address the main challenges of transitioning to renewable energy—namely, reduced system inertia and increased operational complexity—by providing an end-to-end framework for unified grid planning and real-time control.

The algorithm showcases impressive economic efficiency, addressing a major hurdle in smart grid deployment: the exorbitant cost of large-scale flexible alternating current transmission system installations, each priced at \$ 20 million and necessitating years of regulatory approval~\cite{habur2004facts, longoria2022impact}. In addition to its economic advantages, the AAC algorithm exhibits a notable capability: simultaneously stabilizing both the transient regime and the steady state regime. Although frequency fluctuations and power flow deviations appear independent at first glance, their reductions under the AAC intervention are strongly correlated, as shown in Fig.~\ref{fig:4}. This indicates that the AAC algorithm successfully overcomes a challenge posed by system nonlinearity which typically obscures the relationship between these regimes.

While the results demonstrate the effectiveness of the AAC algorithm, several challenges remain to be addressed in future works. First, a simplified version of the UK power grid has been used due to high computational demands. The scalability of the AAC algorithm to full-scale operational grids remains to be validated. Second, the current framework considers only single-line faults. In principle, the GNN-based architecture can accommodate multiple concurrent faults straightforwardly by simultaneously modifying multiple entries of the admittance matrix. However, the number of possible fault configurations grows combinatorially, making the required training computationally demanding with current resources. Extending the framework to address multiple concurrent faults therefore remains an important direction for future work. Third, the AAC algorithm has not yet been fully validated with existing grid dispatching systems. Developing compatible interface protocols would be important for practical deployment.

As power grids experience their most significant change in 100 years, AI offers a viable route to ensure stability. Beyond power grids, this work suggests broader potential applications within network science to suppress cascading failures through adaptive link weight control. The failure mechanism we address—where a single disruption triggers flow redistribution that overloads other links—appears across critical infrastructures, including road closures causing congestion cascades, fuse failures triggering electrical outages, and node failures redirecting data flow. The core insight of the AAC framework—that controlling a small number of critical links can prevent system-wide cascades—may provide a general template for domains in which nonlinear dynamics render analytical prediction intractable. Our findings indicate that intelligent, adaptive control is essential for reliable power grid operation and could prove valuable for other complex networked systems.

\begin{figure*}[ht]
\centering
\includegraphics[width=0.99\linewidth]{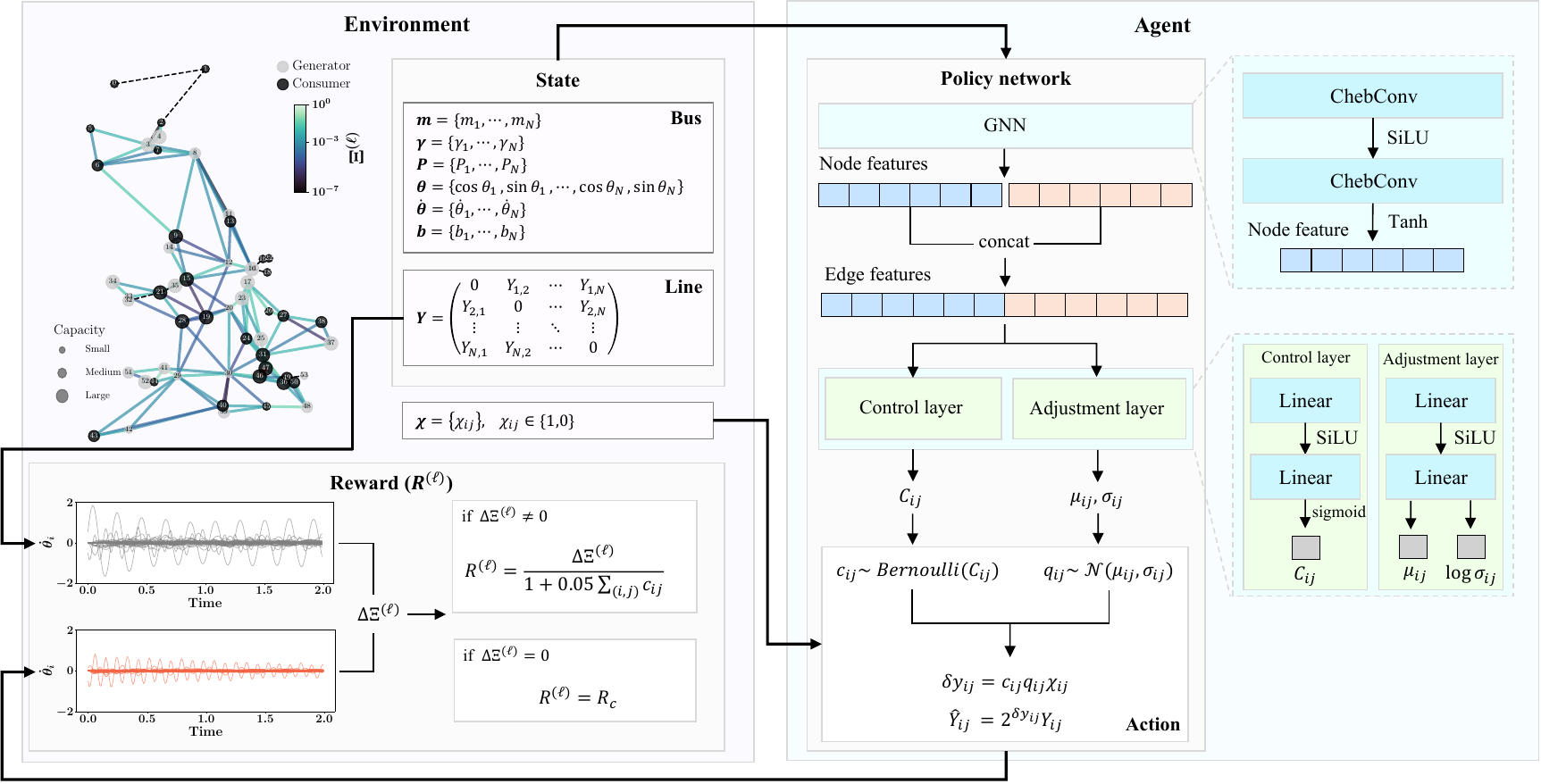}
\caption{Reinforcement learning framework for the AAC algorithm.
The agent observes the power grid state, including system parameters and failed transmission line $\ell$. Based on this observation, it outputs two control signals for each non-failed line: $C_{ij}$ from control layer and $\mu_{ij}, \sigma_{ij}$ from adjustment layer. A binary decision $c_{ij} \sim Bernoulli(C_{ij})$ indicating whether to apply control, and a continuous adjustment magnitude $q_{ij} \sim \mathcal{N}(\mu_{ij}, \sigma_{ij})$. These are combined with regulator presence vector $\chi_{ij}$ into the final action $\delta y_{ij}$. The admittances in power grid are then updated to $\hat{Y}_{ij}$ and a reward $R^{(\ell)}$ is evaluated, which reflects both frequency stability and control efficiency. The AAC is trained to maximize the averaged reward over all possible $\ell$ by iterating this loop.}
\label{fig:5}
\end{figure*}

\section{Methods} \label{sec:method}
We formulate the stability control problem as an RL task in which an agent learns to adjust transmission line admittances after faults. The task requires making two interdependent choices: which lines to control and to what extent. This results in a high-dimensional search space, which increases exponentially with the number of regulators.

\textbf{Single-step episode.}
A standard RL setting expects a sequential Markov Decision Process where the agent takes multiple actions in response to the environment. However, in practical grid operations, repeated interventions increase operational cost. To reflect this, we adopt a single-step episode design in which the agent takes action only once per fault event.
Fig.~\ref{fig:5} illustrates the general framework of the AAC algorithm and its interaction with the environment: the power grid.

\textbf{Positioning of the regulator.}
The regulators are assumed to be located on all transmission lines.
Exceptionally, to obtain Fig.~\ref{fig:3}, some regulators are removed based on their ranking.
Note that the AAC result for the top 5 regulators is trained from scratch under these conditions.

\textbf{State representation.}
The state of the system $X$ is provided as input to the agent.
It comprises two types of inputs: node features and edge features.
For each node $i$, $m_i, \gamma_i, \cos\theta_i, \sin\theta_i, \dot{\theta}_i, b_i$ are concatenated into a single feature vector, where $b_i$ is a binary value that indicates whether the bus is connected to the fault line.
The admittance matrix of the modified transmission lines $Y_{uv}$ is provided as a edge feature, with $Y_{ij} = 0$ for absent lines, thereby also providing the structural topology of the grid.

\textbf{Policy network and action space.}
To capture long-range interactions, we employ the Chebyshev convolution~\cite{defferrard2016convolutional}, a branch of GNN. It incorporates features from distant nodes to update the target node's feature.
Based on observation $X$, the node features are transformed into high-dimensional features through the convolution layers.
For each transmission line $(i,j)$, the characteristic of the line is constructed by concatenating the node features of $i$ and $j$.
The policy network has three output neurons for the regulator line through separate layers: $C_{ij} \in [0, 1]$ from the control layer, and $\mu_{ij}$ and $\sigma_{ij}$ from the adjustment layer.
From $C_{ij}$, binary control decision $c_{ij} \sim \pi_c= Bernoulli(C_{ij})$ is sampled during training and $c_{ij} = C_{ij}$ during evaluation.
Similarly, from $\mu_{ij}$ and $\sigma_{ij}$, adjustment magnitude $q_{ij} \sim \pi_q = \mathcal{N}(\mu_{ij}, \sigma_{ij})$ is sampled during training and $q_{ij} = \mu_{ij}$ during evaluation.
$c_{ij}$ and $q_{ij}$ are combined with $\chi_{ij} \in \{0, 1\}$, which indicates whether the line $(i,j)$ is equipped with a regulator.
Finally, the adjusted admittance $\hat{\boldsymbol{Y}}$ is obtained as
\begin{equation}
\hat{Y}_{ij} = 2^{\delta y_{ij}} Y_{ij}, \quad \text{where } \delta y_{ij} = c_{ij}q_{ij}\chi_{ij}.
\end{equation}
This formulation allows for both increases ($\delta y_{ij}>0$) and decreases ($\delta y_{ij}<0$) in admittance, with $\hat{Y}_{ij} = Y_{ij}$ when $c_{ij}=0$\\

\textbf{Reward design.}
The environment evaluates the updated state and computes a reward $R^{(\ell)}$ to guide policy learning.
The reward balances stability improvement against control complexity:
$$
R^{(\ell)} =
\begin{cases}
\displaystyle\frac{\Delta\Xi^{(\ell)}}{1+0.05\sum_{(i,j)} c_{ij}} & \text{if } \Delta \Xi^{(\ell)} \neq 0, \\
R_{c} & \text{otherwise}.
\end{cases}
$$
To compute the decrease in frequency fluctuation $\Delta \Xi^{(\ell)}$, we set $T=2$ seconds for rapid policy updates during training, while $T=10$ seconds is used in evaluation to complete the evaluation of long-term stability.
The denominator of $R^{(\ell)}$ penalizes excessive use of regulators, and the coefficient $0.05$ is empirically chosen as the value that yielded the most stable training performance. Smaller values lead to excessive intervention even in low-risk scenarios, while larger values cause the agent to avoid control actions altogether.
The constant $R_c$ provides a small positive reward to discourage unnecessary interventions when control has no effect. During training, we chose $R_c = 10^{-5}$ for the UK power grid, as this value produced the best performance.\\

\textbf{Policy optimization.}
As discussed above, the AAC algorithm has two decision distributions: $\pi_c$ for $c_{ij}$ and $\pi_q$ for $q_{ij}$. The GNN processes the structured input $X$ and produces the Bernoulli parameters $C_{ij}$ for the control policy $\pi_c$ and the Gaussian parameters $(\mu_{ij}, \sigma_{ij})$ for the adjustment policy $\pi_q$. 
The joint probability distribution can be written as
\begin{equation}
    \log \pi(\boldsymbol{c}, \boldsymbol{q}|X)
    = \log \pi_c (\boldsymbol{c}|X) + \boldsymbol{c} \log \pi_q(\boldsymbol{q}|X),
\end{equation}
where $\boldsymbol{c}$ acts as a mask for $\pi_q$, ensuring that the adjustment distribution contributes only when the control is applied.

To find the optimal policy $\pi(\boldsymbol{c}, \boldsymbol{q}|X)$ that maximizes reward $R^{(\ell)}$, we employ Proximal Policy Optimization (PPO)~\cite{schulman2017proximal}.
The standard PPO uses a value network to estimate a baseline for calculating the advantage. However, value network often requires sufficient training and may yield inaccurate advantages when underfitting~\cite{moalla2024no}. Several previous studies, therefore, omit them in similar contexts~\cite{shao2024deepseekmath, rafailov2023direct, zhang2024zeroth}.

In our single-step setting, the reward is immediate, so the value network is not required. Accordingly, the loss function of the AAC algorithm utilizing PPO is defined as
\begin{widetext}
\begin{equation}
\mathcal{L}^{(\ell)}=-\max\Biggl[\frac{\pi(\boldsymbol{c}, \boldsymbol{q}|X)}{\pi^{\rm old}(\boldsymbol{c}, \boldsymbol{q}|X)}R^{(\ell)}, \,
\mathrm{clip}\Bigl(\frac{\pi(\boldsymbol{c}, \boldsymbol{q}|X)}{\pi^{\rm old}(\boldsymbol{c}, \boldsymbol{q}|X)}, 1-\epsilon, 1+\epsilon\Bigr)R^{(\ell)}\Biggr],
\end{equation}
\end{widetext}
where $\pi^{\rm old}(\boldsymbol{c}, \boldsymbol{q}|X)$ denotes previous joint policy, and $\epsilon=0.1$ is chosen for the clipping parameter.

$\mathcal{L}^{(\ell)}$ is averaged on all $\ell$-th single-line faults, and the policy network is updated through gradient ascent to maximize expected reward. The detailed training and inference procedures are summarized in SI Sec.~III.

During the preparation of this work, the author(s) used Claude Sonnet 4.5 and ChatGPT 5.2 to improve the clarity and readability of the language. After using these tool/service, the author(s) reviewed and edited the content as needed and take(s) full responsibility for the content of the publication.

\begin{acknowledgments}
This work was supported by the Korea Institute of Energy Technology Evaluation and Planning(KETEP) and the Ministry of Trade, Industry \& Energy(MOTIE) of the Republic of Korea (No. 20224000000100) (BK), and the National Research Foundation of Korea (NRF) grant funded by the Korea Government (MSIT) (No. RS-2025-00556024) (HY).
\end{acknowledgments}

\section*{Supplementary Information}
\label{seca:Supplementary Information}

\subsection{Topological and Dynamical Characteristics of the SHK Network}
To verify that the performance of the AAC algorithm is not confined to heterogeneous real-world grids, we tested it on an SHK network with more homogeneous parameters. The SHK network is a synthetic model commonly used to investigate the dynamics of the power grid. Its topology is determined by four parameters $p,~q,~r,~\mathrm{and}~s$. In this study, we generate the SHK network whose properties closely match those of the reduced UK power grid. To identify suitable parameters, we compare several topological metrics with those of the UK grid (Fig.~1(a) of the main text): the second-smallest eigenvalue of the Laplacian matrix $\lambda$, mean degree $\bar{k}$, mean clustering coefficient (cc), diameter, and average length of the shortest path (aspl). We find that $p=0.4,~q=0.9,~r=0.1,~\mathrm{and}~s=0.2$ agrees closely across all metrics. Supplementary Table~1 shows that the resulting SHK network matches the UK grid not only in network size $N$ and number of links $L$, but also in its key topological characteristics.

\begin{table*}[ht]
\centering
\renewcommand{\arraystretch}{1.5}
\resizebox{0.8\textwidth}{!}{%
\begin{tabular}{|>{\centering\arraybackslash}m{1.5cm}
                |>{\centering\arraybackslash}m{1.5cm}
                |>{\centering\arraybackslash}m{1.5cm}
                |>{\centering\arraybackslash}m{1.5cm}
                |>{\centering\arraybackslash}m{1.5cm}
                |>{\centering\arraybackslash}m{1.5cm}
                |>{\centering\arraybackslash}m{1.5cm}
                |>{\centering\arraybackslash}m{1.5cm}|}
\hline
    & $N$  & $L$   & $\bar{k}$ & $\lambda$ & cc & diameter & aspl \\ \hline
UK  & 54 & 114 & 4.222 & 0.1212 & 0.5025 & 10 & 3.712  \\ \hline
SHK & 54 & 115 & 4.259 & 0.1187 & 0.4070 & 8 & 3.703  \\ \hline
\end{tabular}
}
\caption{Topological properties of UK power grid and the SHK network ($p=0.4,~q=0.9,~r=0.1,~s=0.2$).}
\end{table*}

In the SHK network, generators and consumers are randomly assigned (Supplementary Fig.~1). To study a relatively homogeneous power grid, we set most of the parameters of the swing equation [Eq.~(1) in the main text] to uniform values: generator and consumer powers are assigned as $+1$ or $-1$, respectively; inertia constants $m_i$ and damping coefficients $\gamma_i$ are set to the average values observed in the UK power grid; and all coupling strengths are fixed at $K_{ij} = 4$. As shown in Supplementary Fig.~1, even with reduced parameter heterogeneity, the frequency fluctuations $\Xi^{(\ell)}$ over 10 seconds still depend on the location of the fault line $\ell$. Therefore, in a relatively homogeneous grid, effective regulator locations must still be tailored to each specific fault scenario.

\begin{figure}[ht]
    \includegraphics[width=0.5\linewidth]{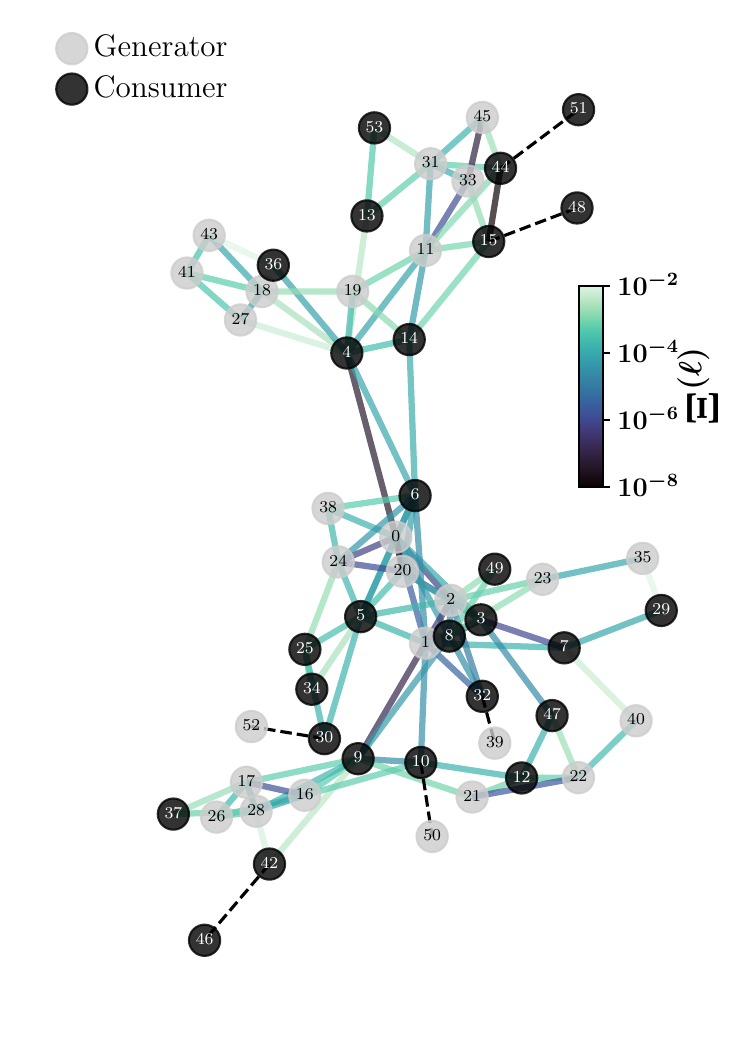}
    \caption{The frequency fluctuations $\Xi^{(\ell)}$ in the generated SHK network ($p=0.4,~q=0.9,~r=0.1,~s=0.2$). Black dashed lines indicate fault scenarios that are excluded from the analysis, as disconnecting these lines would split the power grid into two isolated parts.}
    \label{fig:shk_network}
\end{figure}

\subsection{AAC algorithm Performance in the SHK Network}
In the main text, we showed that the adaptive admittance controller (AAC) algorithm can effectively decrease $\Xi^{(\ell)}$ in the UK power grid, which exhibits heterogeneity in the parameters of the swing equation [Eq.~(1)] of the main text. Here, we analyze its performance in the SHK network, which has more homogeneous characteristics than the UK grid. To ensure effective performance, it is necessary to choose an appropriate $R_c$, as this parameter influences the control tendencies of the AAC algorithm. We find that $R_c = 10^{-5}$ produces appropriate behavior. As shown in Supplementary Fig.~2, the AAC algorithm reduces $\Xi^{(\ell)}$ in high-impact fault scenarios while maintaining $\Xi^{(\ell)}$ in low-impact scenarios by avoiding unnecessary interventions. The AAC algorithm achieves an approximately 55\% reduction in the average frequency fluctuation in the SHK network. This shows that the AAC algorithm can effectively suppress $\Xi^{(\ell)}$ even in a homogeneous power grid.

\begin{figure}[ht]
    \includegraphics[width=0.9\linewidth]{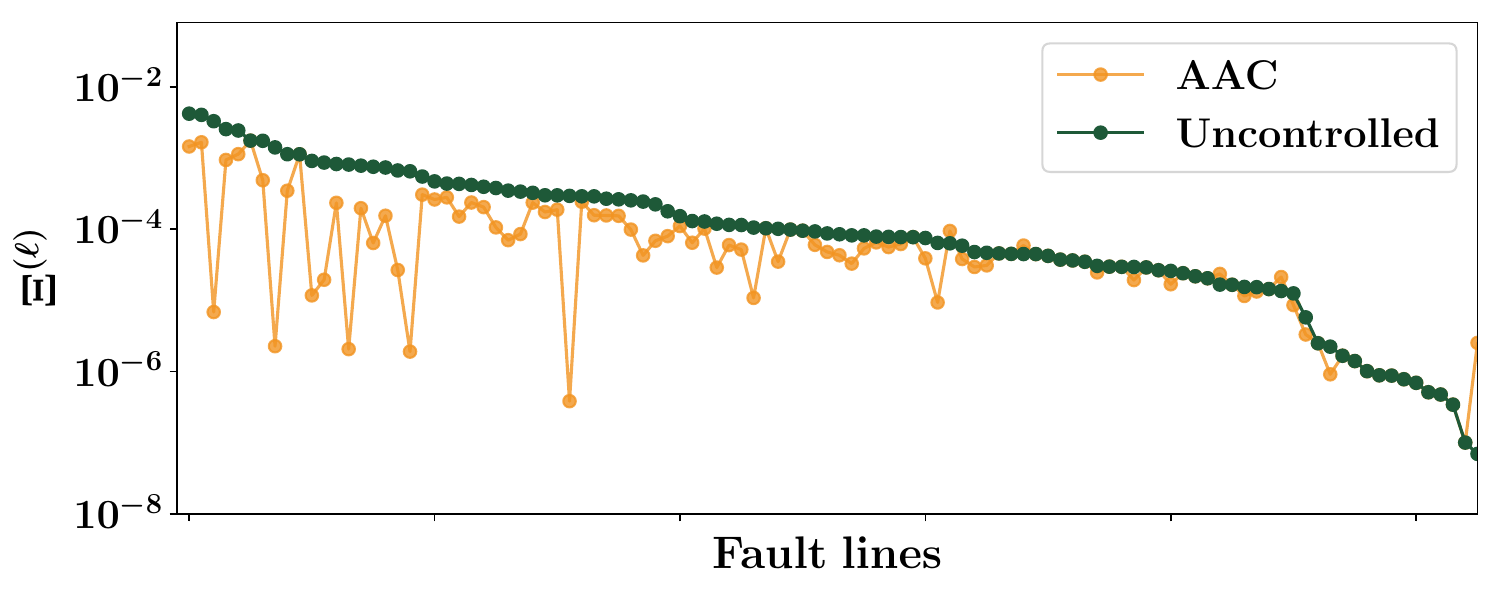}
    \caption{The frequency fluctuations $\Xi^{(\ell)}$ for the non-control case (green) and the AAC algorithm case (orange) across all single-line faults in the SHK network. Fault scenarios are sorted by the magnitude of $\Xi^{(\ell)}$ in the uncontrolled case. AAC algorithm results show that $\Xi^{(\ell)}$ is significantly reduced in high-$\Xi^{(\ell)}$ scenarios, while in low-$\Xi^{(\ell)}$ cases the algorithm tends to avoid intervention.}
    \label{fig:shk_fig2}
\end{figure}

\begin{figure}[ht]
    \includegraphics[width=0.9\linewidth]{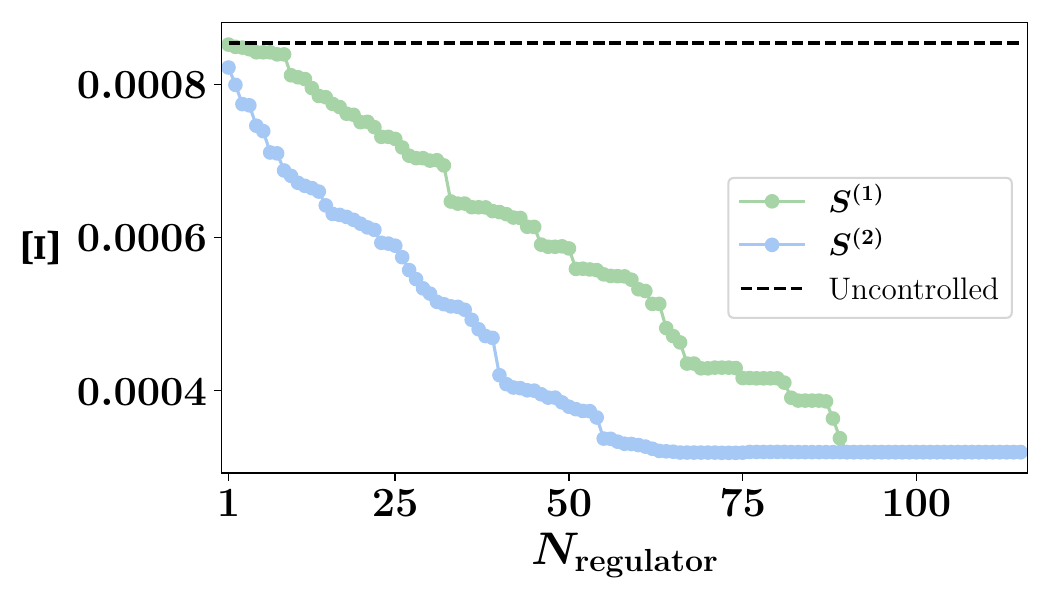}
    \caption{Averaged frequency fluctuation $\Xi$ vs. the number of regulators $N_{\rm regulator}$ installed on the transmission lines in the SHK network. The regulators are ordered by rank for each metric. $S^{(2)}$ achieves the best performance by only using less than one-half of all regulators. 
    }
    \label{fig:shk_fig3}
\end{figure}

\begin{figure}[ht]
    \includegraphics[width=0.9\linewidth]{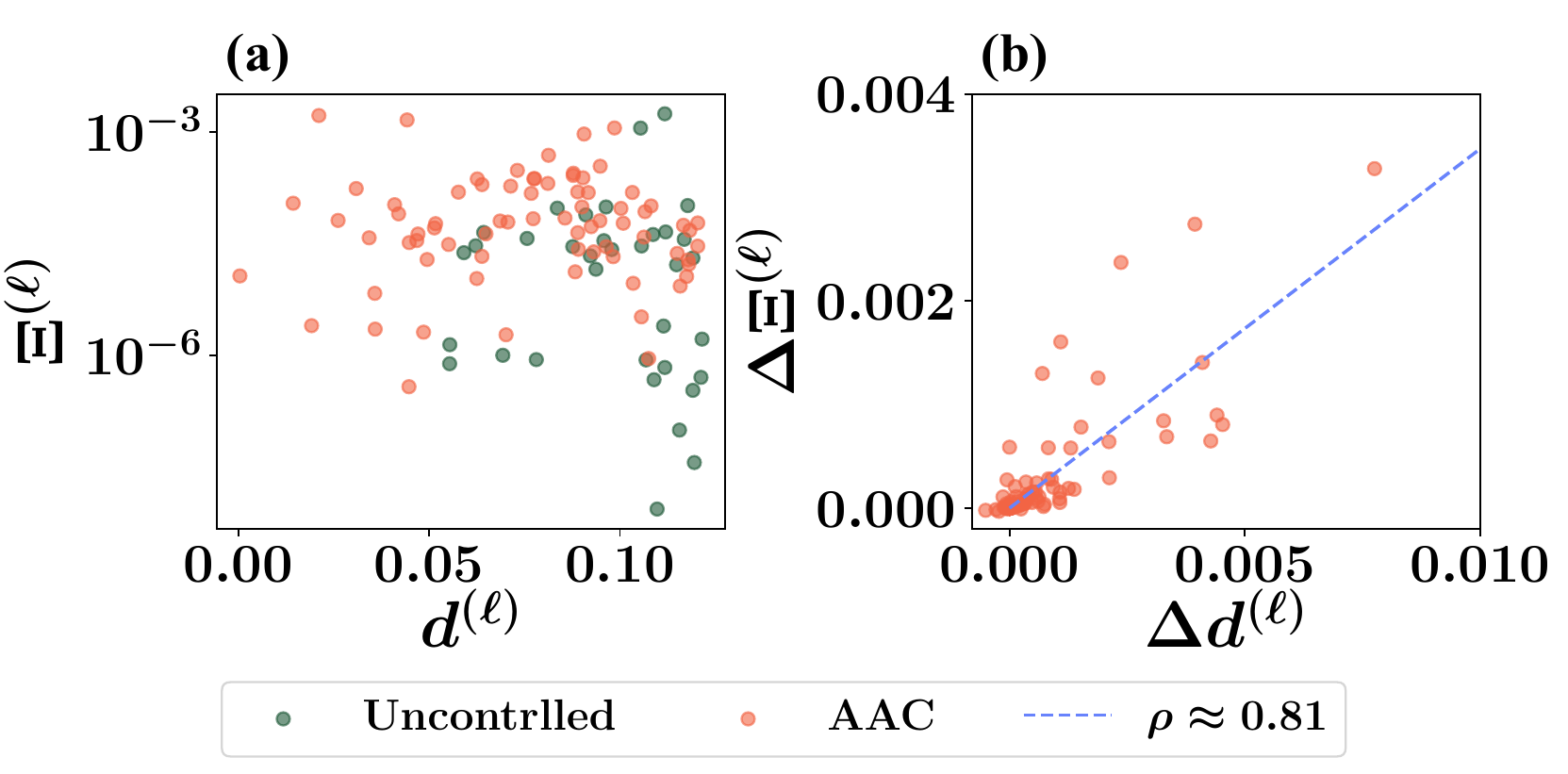}
    \caption{Phase space analysis of the AAC algorithm's operating mechanism in the SHK network. (a) Scatter plot showing the relationship between frequency fluctuations $\Xi^{(\ell)}$ and power flow variation $d^{(\ell)}$. The two quantities exhibit largely independent behavior. (b) The scatter plot shows the relationship between the reduction in frequency fluctuations $\Delta \Xi^{(\ell)}$ and power flow restoration $\Delta d^{(\ell)}$. Each point corresponds to $\ell$-th single-line fault in the SHK network.}
    \label{fig:shk_fig4}
\end{figure}

The AAC algorithm adjusts $Y_{ij}$ differently depending on the fault scenario, resulting in certain locations being frequently modified (Fig.~2 of the main text). A similar pattern is observed in the SHK network. Supplementary Fig.~3 shows the average $\Xi$ across all fault scenarios as a function of the number of regulators $N_{\rm regulator}$ used. Unlike in the UK power grid, we omit PTDF-based measurements due to the absence of a slack bus, and $S^{(3)}$ because all $K_{ij} = 4$ in the SHK network, making $S^{(3)}$ equivalent to $S^{(2)}$ up to a constant factor.

As shown in Supplementary Fig.~3, the minimum fluctuation level can be achieved without installing regulators on all transmission lines. In particular, $S^{(2)}$ reaches the minimum $\Xi$ with only 74 regulators out of 115 possible lines, demonstrating that strategic placement on a subset of lines is sufficient for optimal performance. $S^{(2)}$ achieves the best performance compared to $S^{(1)}$, confirming that $S^{(2)}$ is the most effective ranking metric for both the SHK network and the UK grid.

Although the AAC algorithm effectively suppresses frequency fluctuations in the transient regime (Supplementary Fig.~2), the steady state stability is not guaranteed due to the system's nonlinearity. To evaluate this, we examine the power flow variation $d^{(\ell)}$ in phase space, as defined in Eq.~(4) of the main text.

As shown in Supplementary Fig.~4(a), $\Xi^{(\ell)}$ and $d^{(\ell)}$ exhibit negligible correlation, indicating that transient regime frequency fluctuations and power flow deviations in the steady state are largely independent due to nonlinearity. In contrast, Supplementary Fig.~4(b) reveals that control-induced changes are strongly correlated. The Pearson correlation coefficient $\rho$ between $\Delta \Xi^{(\ell)}$ and $\Delta d^{(\ell)}$ is approximately 0.81, demonstrating that the AAC algorithm effectively bridges the nonlinearity-induced gap between the two regimes. These results confirm that the AAC algorithm simultaneously stabilizes both transient dynamics and steady state in the SHK network.

\subsection{Detailed Procedures of the AAC Algorithm}
Algorithms 1 and 2 illustrate the detailed procedures of the Adaptive Admittance Controller (AAC) algorithm. Here, $\phi$ denotes the trainable parameters of the policy network, $\mathcal{F}$ is the set of all single-line fault scenarios, and $B$ is the batch buffer that stores transitions. $\pi_{\text{old}}$ denotes the policy before the current update. The policy is updated over $N_{\text{epoch}}$ epochs using the learning rate $\eta$.

\begin{algorithm}[ht]
\caption{AAC Training}
\label{alg:aac_training}

Initialize policy network parameters $\phi$ randomly\;

\For{\rm epoch $=1$ \KwTo $N_{\text{epoch}}$}{
    Initialize batch buffer $B \leftarrow \emptyset$\;

    \For{\rm each fault scenario $\ell=(u,v)\in\mathcal{F}$}{
        \textit{\textbf{State Construction:}}\;
        Set $Y_{uv}\leftarrow 0$, $Y_{vu}\leftarrow 0$\tcp*{Apply line fault}
        Construct state
        $X \leftarrow \{\boldsymbol{m},\boldsymbol{\gamma},\boldsymbol{P},
        \boldsymbol{\theta},\dot{\boldsymbol{\theta}},
        \boldsymbol{b},\boldsymbol{Y}\}$\;

        \BlankLine
        \textit{\textbf{Policy Network Forward Process:}}\;
        $\boldsymbol{H}\leftarrow \text{GNN}(X,\boldsymbol{Y})$\;

        \For{\rm each non-failed line $(i,j)\in E\setminus\{\ell\}$}{
            $\boldsymbol{h}_{ij}\leftarrow \text{Concat}(H_i,H_j)$\;
            $C_{ij}\leftarrow
            \text{Sigmoid}(\text{MLP}_{\text{control}}(\boldsymbol{h}_{ij}))$\;
            $\mu_{ij},\log\sigma_{ij}
            \leftarrow \text{MLP}_{\text{adjust}}(\boldsymbol{h}_{ij})$\;
        }

        \BlankLine
        \textit{\textbf{Action Sampling:}}\;
        \For{\rm each non-failed line $(i,j)\in E\setminus\{\ell\}$}{
            $c_{ij}\sim \text{Bernoulli}(C_{ij})$\;
            $q_{ij}\sim \mathcal{N}(\mu_{ij},\exp(\log\sigma_{ij}))$\;
            $\delta y_{ij}\leftarrow c_{ij}\cdot q_{ij}\cdot\chi_{ij}$\;
            $\hat{Y}_{ij}\leftarrow 2^{\delta y_{ij}}\cdot Y_{ij}$\;
        }

        \BlankLine
        \textit{\textbf{Environment Evaluation:}}\;
        Apply $\hat{\boldsymbol{Y}}$ to grid and simulate swing equation for $T_{\text{train}}$\;
        Compute $\Xi^{(\ell)}_{\text{ctrl}}$ and $\Xi^{(\ell)}_{\text{unctrl}}$ via Eq.~(2)\;
        $\Delta\Xi^{(\ell)}
        \leftarrow
        \Xi^{(\ell)}_{\text{unctrl}}
        -
        \Xi^{(\ell)}_{\text{ctrl}}$\;

        \BlankLine
        \textit{\textbf{Reward Computation:}}\;
        \eIf{$\Delta\Xi^{(\ell)}\neq 0$}{
            $R^{(\ell)}
            \leftarrow
            \Delta\Xi^{(\ell)}
            /
            \bigl(1+0.05\sum_{(i,j)}c_{ij}\bigr)$\;
        }{
            $R^{(\ell)}\leftarrow R_c$\;
        }

        Compute $\log\pi_\phi(\boldsymbol{c},\boldsymbol{q}\mid X)$ via Eq.~(6)\;
        Append $(X,\boldsymbol{c},\boldsymbol{q},R^{(\ell)},\log\pi_\phi)$ to $B$\;
    }

    \BlankLine
    \textit{\textbf{PPO Policy Update:}}\;
    Store $\pi_{\text{old}}\leftarrow\pi_\phi$\;

    \For{\rm each $(X,\boldsymbol{c},\boldsymbol{q},R^{(\ell)},\log\pi_{\text{old}})\in B$}{
        $\mathcal{L}^{(\ell)}
        \leftarrow
        -\max\!\bigl[
        (\pi_\phi/\pi_{\rm old})R^{(\ell)},
        \text{clip}(\pi_\phi/\pi_{\rm old},1-\epsilon,1+\epsilon)R^{(\ell)}
        \bigr]$\;
    }

    $\mathcal{L}
    \leftarrow
    \frac{1}{|\mathcal{F}|}\sum_\ell \mathcal{L}^{(\ell)}$\;
    $\phi\leftarrow\phi-\eta\nabla_\phi\mathcal{L}$\;
}

\KwRet{$\pi_\phi$}\;

\end{algorithm}

\begin{algorithm}[ht]
\caption{AAC Evaluation (Inference)}
\label{alg:aac_eval}

\KwIn{Trained policy $\pi_\phi$, fault scenario $\ell=(u,v)$}
\KwOut{Adjusted admittance matrix $\hat{\boldsymbol{Y}}$}

Set $Y_{uv}\leftarrow 0$, $Y_{vu}\leftarrow 0$\;

Construct state
$X \leftarrow
\{\boldsymbol{m},\boldsymbol{\gamma},\boldsymbol{P},
\boldsymbol{\theta},\dot{\boldsymbol{\theta}},
\boldsymbol{b},\boldsymbol{Y}\}$\;

$\boldsymbol{H}\leftarrow \text{GNN}(X,\boldsymbol{Y})$\;

\For {\rm each non-failed line $(i,j)\in E\setminus\{\ell\}$}{
    $\boldsymbol{h}_{ij}\leftarrow \text{Concat}(H_i,H_j)$\;
    $C_{ij}\leftarrow
    \text{Sigmoid}(\text{MLP}_{\text{control}}(\boldsymbol{h}_{ij}))$\;
    $\mu_{ij}\leftarrow\text{MLP}_{\text{adjust}}(\boldsymbol{h}_{ij})$\;
    $c_{ij}\leftarrow C_{ij}$\tcp*{Deterministic}
    $q_{ij}\leftarrow \mu_{ij}$\tcp*{Deterministic}
    $\hat{Y}_{ij}
    \leftarrow
    2^{c_{ij}\cdot q_{ij}\cdot\chi_{ij}}\cdot Y_{ij}$\;
}

Apply $\hat{\boldsymbol{Y}}$ and simulate swing equation for $T_{\text{eval}}$\;

\KwRet{$\hat{\boldsymbol{Y}}$}\;

\end{algorithm}

\end{document}